\documentclass[prx,aps,reprint,noshowpacs,superscriptaddress,floatfix,letterpaper,longbibliography]{revtex4-2}
\usepackage{amsmath,amssymb,amsbsy,amsfonts,amsthm,bbm,bm,mathtools,mathrsfs}
\usepackage{color}
\usepackage{physics}
\usepackage{xfrac}
\usepackage[dvipsnames]{xcolor}
\definecolor{LapisLazuli}{RGB}{47, 102, 169}
\usepackage[colorlinks=true,citecolor=LapisLazuli,linkcolor=LapisLazuli,urlcolor=LapisLazuli]{hyperref}
\usepackage{empheq}
\usepackage{pgfplots}
\usepackage{stackengine}
\usepackage{relsize}
\usepackage[inline]{enumitem}
\usepackage[normalem]{ulem}
\usepackage{comment}
\usepackage{import}
\usepackage[english]{babel}
\UseRawInputEncoding
\usepackage{pgfplots}
\pgfplotsset{compat = newest}
\usetikzlibrary{arrows,intersections}
\usepackage{tikz-3dplot}
\usepackage{tikz}

\usepackage{mathrsfs}
\usepackage[mathscr]{euscript}
\usepackage{soul}

\usepackage{calligra}
\DeclareMathAlphabet{\mathcalligra}{T1}{calligra}{m}{n}
\DeclareFontShape{T1}{calligra}{m}{n}{<->s*[2.2]callig15}{}

\newcommand{\bx}{\boldsymbol{x}}
\newcommand{\bq}{\boldsymbol{q}}
\newcommand{\bp}{\boldsymbol{p}}
\newcommand{\stability}{{\bm{A}}}

\newcommand{\yu}{{\bm{u}}}
\newcommand{\bphi}{{\bm{\phi}}}
\newcommand{\brho}{{\bm{\varrho}}}
\newcommand{\bxi}{{\bm{\xi}}}

\graphicspath{{../plots/}} 

\usepackage{pgfplots}
\pgfplotsset{compat = newest}
\usetikzlibrary{arrows,intersections}
\usepackage{tikz-3dplot}
\usepackage{tikz}

\makeatletter
\def\maketitle{
	\@author@finish
	\title@column\titleblock@produce
	\suppressfloats[t]}
\makeatother

\begin{document}

\title{Maximum speed of dissipation} 

\author{Swetamber~Das}
\author{Jason~R.~Green}
\email[]{jason.green@umb.edu}
\affiliation{Department of Chemistry,\
	University of Massachusetts Boston,\
	Boston, Massachusetts 02125, USA
}
\affiliation{Department of Physics,\
	University of Massachusetts Boston,\
	Boston, Massachusetts 02125, USA
}

\date{\today}

\begin{abstract}

We derive statistical-mechanical speed limits on dissipation from the classical, chaotic dynamics of many-particle systems.
In one, the rate of irreversible entropy production in the environment is the maximum speed of a deterministic system out of equilibrium, $\bar S_e/k_B\geq 1/2\Delta t$, and its inverse is the minimum time to execute the process, $\Delta t\geq k_B/2\bar S_e$.
Starting with deterministic fluctuation theorems, we show there is a corresponding class of speed limits for physical observables measuring dissipation rates.
For example, in many-particle systems interacting with a deterministic thermostat, there is a trade-off between the time to evolve between states and the heat flux, $\bar{Q}\Delta t\geq k_BT/2$.
These bounds constrain the relationship between dissipation and time during nonstationary process, including transient excursions from steady states.

\end{abstract}

\maketitle
\textit{Introduction.---} When a piece of macroscopic matter is set in motion, it will irreversibly lose energy due to frictional or viscous heat generation~\cite{Callen1985}.
If the system undergoes these dissipative transitions on finite timescales, it will expel energy as heat, wasting free energy and producing entropy.
But, how quickly does a, potentially finite-size, classical system of particles incur these thermodynamic costs as it evolves and interacts with its environment?
Are there limits on the speed at which these observables accumulate during finite time processes?
Answering these questions is important to both synthetic and biological systems.
For example, the microscopic machinery of the cell includes nanoscale motors, such as kinesin, that perform essential functions on finely-tuned timescales~\cite{Goel2008}.
Given one second to power motion and haul molecular cargo, kinesin will dissipate about $650k_BT$ of energy to its environment~\cite{Busta2005}, where $k_B$ is Boltzmann's constant and $T$ is the temperature. 
One might then hypothesize that the dissipation of energy and the production of entropy could constrain the time it takes for such systems to transition between states~\cite{BryantMachta2020, Skinner2021}.

Recently, there have been stochastic thermodynamic predictions of the speed of irreversible processes driven by gradients in temperature, pressure, and concentration.
These thermodynamic speed limits~\cite{shiraishiSpeedLimitClassical2018,hasegawaUncertaintyRelationsStochastic2019,itoStochasticTimeEvolution2020,nicholsonNonequilibriumUncertaintyPrinciple2018,FalEsp2020,nicholsonTimeInformationUncertainty2020} are giving new understanding of the trade-off between time and dissipation.
In some forms, these purely stochastic results are analogues of quantum speed limits~\cite{Deffner_2017}, which bound the speed at which physical systems (or their observables) evolve between two distinguishable states.
While thermodynamic speed limits on fluxes of energy and entropy are an emerging feature of stochastic thermodynamics, another important feature is fluctuation theorems. However, fluctuation theorems were originally derived for mechanical systems out of equilibrium~\cite{EvansSearles2002} with the techniques of dynamical systems theory~\cite{gaspardChaosScatteringStatistical1998,dorfmanIntroductionChaosNonequilibrium1999}. 
Still open is the question of whether thermodynamic speed limits on dissipation have counterparts that derive from the physical dynamics of mechanical systems.
That is, despite progress in classical speed limits and the classical limit of quantum speed limits~\cite{shanahanQuantumSpeedLimits2018,okuyamaQuantumSpeedLimit2018, Poggi2021, DasGreen2023speed}, there are not speed limits on dissipation that derive from atomistic dynamics and apply to many observables.

In this Letter, we derive a classical speed limit with entropy production, dissipative flux, energy dissipated as heat, and transport coefficients for dissipative, continuous-time dynamical systems.
We show the mean entropy production puts a bound on the time for a classical many-body system to transition between two nonequilibrium states as defined by the integrated deterministic fluctuation theorem.
This trade-off between the time to evolve between two distinguishable states and the entropy production predicts how much entropy or energy will be dissipated by a physical process over a finite period of time or how much time will it take for a process to dissipate an amount of entropy or energy.
Like early fluctuation theorems~\cite{Evans1994,Gallavotti1995,Cohen1999}, this bound can be expressed in terms of Lyapunov exponents, the fundamental quantities characterizing deterministic chaos.
These quantities have been used to analyze rare trajectories~\cite{TailleurK07}, jamming~\cite{banigan_chaotic_2013}, nonequilibrium self-assembly~\cite{GreenCGS13}, equilibrium and nonequilibrium fluids~\cite{evans1990statistical, Bosetti_2014,das_self-averaging_2017}, and critical phenomena~\cite{dascritical2019}.
The finite system size, finite-time behaviors in these examples are subject to speed limits on dissipation.

\begin{figure}[t!]
	\hspace*{.4cm}\includegraphics[width=0.7\columnwidth]{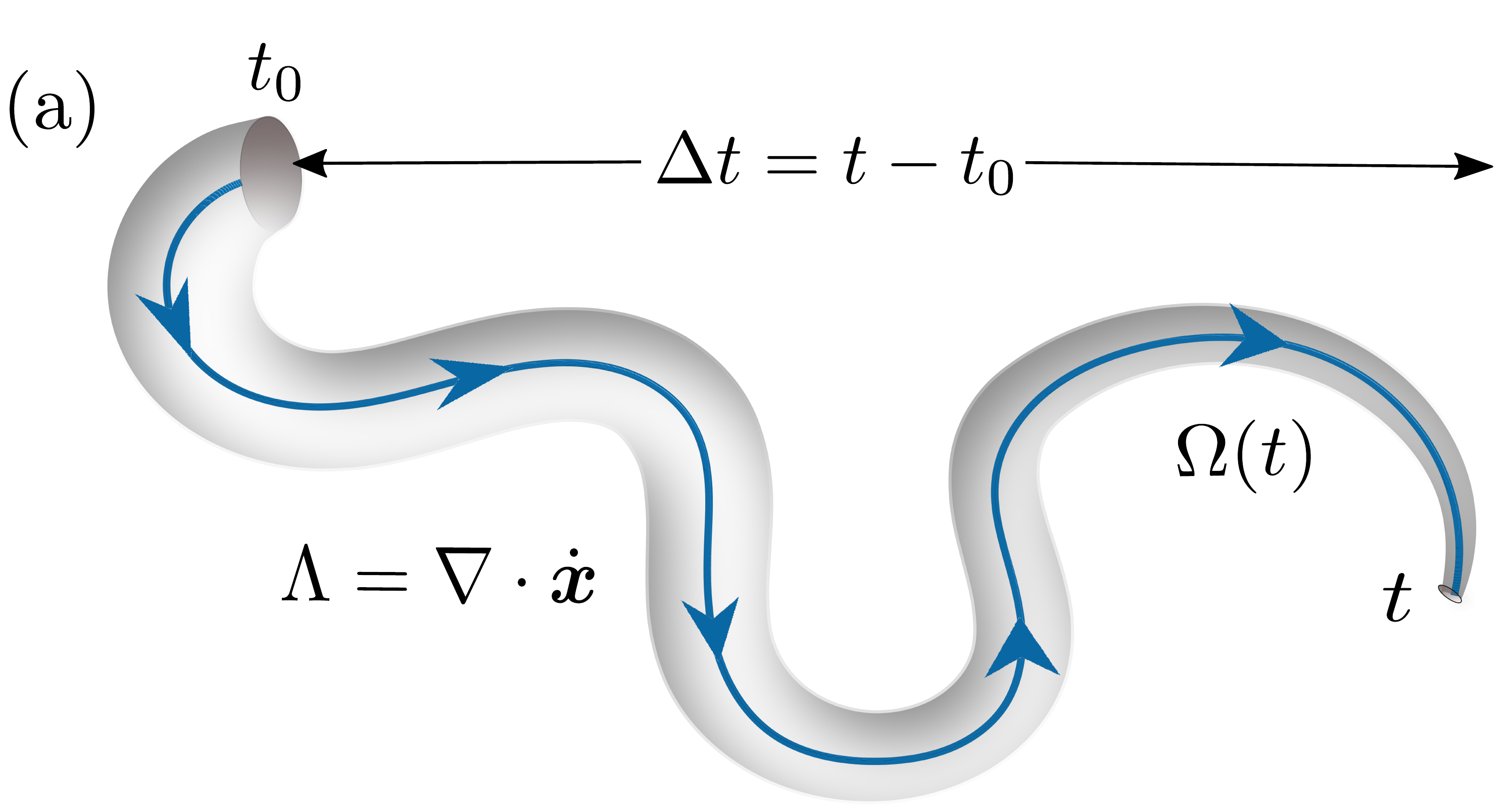}\\
	\vspace*{0.0cm}
	\includegraphics[width=0.95\columnwidth]{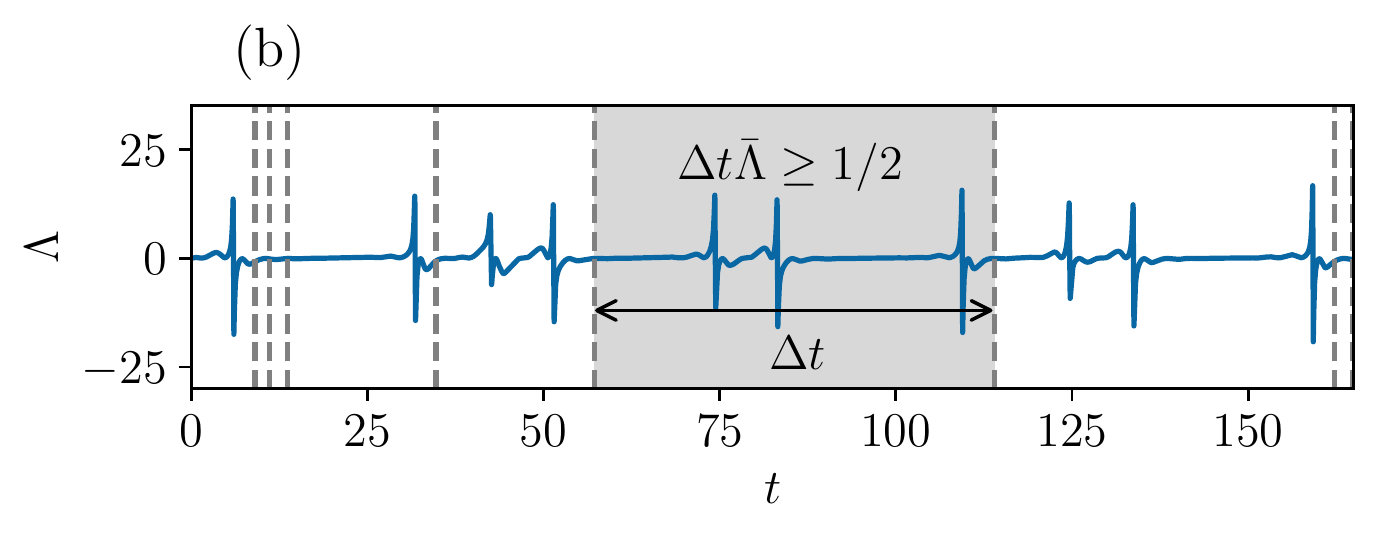}
	\caption{(a) A typical trajectory $\Omega$ evolving in the $n$-dimensional phase space of a dissipative system from an initial time $t_0$ to a final $t$.
An infinitesimal volume surrounding the initial condition contracts at a rate $\bar{\Lambda}$ over the time interval $\Delta t = t - t_0$.
The contraction rate caused by dissipation satisfies the inequality $\Delta t\bar{\Lambda}\geq (1-\eta)/(1+\eta)$ (see text for details). (b) Time evolution of the contraction rate $\Lambda$ at $\epsilon = 0.5$ of a heat-conducting oscillator, the 0532 model~\cite{Hoover2016b,Patra2016}. 
A nonzero value of $\epsilon$ imposes a temperature gradient and the system relaxes to a nonequilibrium steady state.
The gray dashed lines mark time windows that follow the speed limit $\Delta t \bar{\Lambda} \geq 1/2$ for $\eta = 1/3$  (Eq.~\ref{eq:diss_time1}).
The largest time window $\Delta t$ in the time series is highlighted.
 \label{fig:traj-volume-schematic}}
\end{figure}

\textit{Dissipation from phase space volume contraction.---} 
Equilibrium statistical physics has successful methods for counting indistinguishable microscopic states for a given set of physical conditions~\cite{Callen1985}.
There, the thermodynamic entropy has long been a measure of the number of states, $S=k_B\ln \mathcal{V}$ using phase space volume as a surrogate for the raw count.
However, away from equilibrium, dissipative forces cause the mechanical contraction of phase space~\cite{gaspardChaosScatteringStatistical1998,dorfmanIntroductionChaosNonequilibrium1999}.
To construct measures of distinguishability between physical states along deterministic trajectories away from equilibrium, we can take a complementary approach using the relationship between the dissipation of energy and the contraction rate of the phase space volume $\mathcal{V}$~\cite{gaspardChaosScatteringStatistical1998,dorfmanIntroductionChaosNonequilibrium1999}. 
With this approach, we can derive speed limits on dissipation in nonequilibrium processes by measuring the number of distinct physical states visited by a classical trajectory during a time $\Delta t=t-t_0$.

The rate of phase space contraction is a theoretical link between dynamics and distinguishable states that are necessary for speed limits on physical observables.
Consider a $d$-dimensional system of $N$ classical particles at a point, $\bx\in \mathcal{M}$, in phase space $\mathcal{M}$.
As the system evolves under chosen physical conditions, the state traces a trajectory, $\Omega=\{\bx(t); t_0\leq t\leq t_f\}$, governed by the equations of motion $\dot{\bx} = \boldsymbol{F}[\bx(t)]$ in an $n = 2dN$-dimensional phase space, $\mathcal{M}$, Fig.~\ref{fig:traj-volume-schematic}(a).

Infinitesimal phase space volumes, $\delta\mathcal{V}$, surrounding the phase point have an intrinsic rate at which they deform -- the divergence of the flow, $\Lambda := \delta \dot{\mathcal{V}}/\delta \mathcal{V}=\grad\cdot \dot{\boldsymbol{x}}=\sum^{2dN}_{i} \partial F_{i}/\partial x_{i}$. 
The dynamics are dissipative if the time averaged $\Lambda$ is negative in the long-time limit~\cite{Ramshaw2017}.
The instantaneous contraction rate $\Lambda$ is readily calculated for biological and engineered dynamical systems~\cite{PikovskyP16} as well as myriad (bio)molecular systems simulated with molecular dynamics~\cite{evans_morriss_2008}.
When these systems interact with a deterministic thermostat~\cite{Hoover1999}, the phase space contraction rate gives the thermodynamic entropy production rate and transport coefficients~\cite{gaspardChaosScatteringStatistical1998,dorfmanIntroductionChaosNonequilibrium1999}.

When exchanging energy with its surroundings, the deformation of phase space volumes also determines the time evolution of the probability density over phase space $\rho(\bx,t)$.
The density evolves according to Liouville's equation $d_t\rho = \partial_t\rho + \dot{\boldsymbol{x}}\nabla \rho = -\rho\Lambda$.
However, if the number of members of the ensemble is constant, then the equation of motion becomes $d_t\delta\mathcal{V}=+\Lambda\delta\mathcal{V}$ with the solution $\delta\mathcal{V}(t) = \delta\mathcal{V}(t_0)e^{+\int_{t_0}^{t}\Lambda(t') dt'}$.
As a trajectory evolves from an initial state ($\boldsymbol{x}(t_0)$, $\delta \mathcal{V}(t_0)$) to a final state ($\boldsymbol{x}(t)$, $\delta \mathcal{V}(t)$), the infinitesimal volumes are measure of their distinguishability through the compression factor,
\begin{align}
\label{eq:survival}
{\Gamma(t,t_0)}:= 
 e^{2\int_{t_0}^{t}\Lambda[\bm x(t')] dt'}.
\end{align}
We use the square volumes because they are numerically computable as the determinant of classical density matrices~\cite{DasGreen2022} (SM~\ref{SM:EOM-xi}, \ref{SM:geometric-xi}).
While the rate $\Lambda$ at given phase point does not generally have a definite sign, there are finite time intervals $\Delta t = t - t_0$ in which the \textit{net} contraction $\int_{t_0}^{t}\,\Lambda(t') dt'$ is negative if the system is dissipative.

\textit{Classical speed limit from phase space contraction.---} 
Choosing a value of the compression factor defines a criterion that we can use for a speed limit on the contraction of phase space.
Say the dynamics over a time interval $\Delta t$ meet the condition $\Gamma\equiv \eta$ by contracting local volume, $\int_{t_0}^t\Lambda dt' < 0$.
Applying a bilinear approximation~\footnote{The bilinear approximation of $e^{2x}$ leads to the inequality $e^{2x}\geq (1+x)/(1-x)$ for $x\leq 0$.} to Eq.~\ref{eq:survival} gives
\begin{align}\label{eq:diss_time1}
  \Delta t_{\eta} \bar{\Lambda}\geq \frac{1-\eta}{1 + \eta} =: f(\eta) \quad \quad (\bar{\Lambda} > 0, 0<\eta \leq 1),
\end{align}
a trade-off between time and the time average phase space contraction rate, $\bar{\Lambda} = -\Delta t^{-1}\int_{t_0}^{t}\Lambda dt$~\footnote{A similar inequality exists for net \textit{expanding} phase volumes, $\bar{\Lambda} \leq 0$ (SM~\ref{sm-speed-limit}).}. 
This inequality applies to any dissipative, continuous time dynamical system.
To illustrate, Fig.~\ref{fig:traj-volume-schematic}(b) shows $\Lambda$ as a function of time for a singly thermostatted oscillator for $\eta = 1/3$ with intervals marking time windows that satisfy this speed limit.

This inequality has three immediate interpretations:
\smallskip

(1) Rearranging to $\bar \Lambda/f(\eta)\geq \Delta t_{\eta}^{-1}$, we see that the dissipation rate $\bar \Lambda$ is the maximum speed $\Delta t_{\eta}^{-1}$ at which the system can transition between the distinct initial and final states.
It is also the maximum number of distinguishable states the system can pass through per unit time.
That is, it sets a speed limit on classical dynamical systems transitioning between any two $\eta$-distinguishable states.
Alternatively, if the dissipate rate $\bar\Lambda$ is unknown, the time interval can be seen as a lower bound.

(2) This inequality is a classical analogue of recent extensions to the Margolus-Levitin speed limit on quantum dynamics~\footnote{A weaker form of the inequality in Eq.~\ref{eq:diss_time1} follows from the linear approximation, $e^x\geq 1+x$, which gives $\Delta t \bar{\Lambda}/n\geq 1/2$ and a factor of two that one might expect comparing to the Margolus-Levitin speed limit.
For a systems with a time-independent $\Lambda$ and constant rate of phase space contraction, such as the damped harmonic oscillator and the Lorenz-Fetter model, the inequality also simplifies to $-\Delta t\Lambda \geq 1$.}.
Quantum mechanical systems with a finite average energy cannot evolve between two states distinguishable by a fidelity $\epsilon$ in a time shorter than $\Delta t_\perp \geq h\alpha(\epsilon)/4\langle E\rangle$~\cite{margolus1998maximum}, where $\alpha$ is the fidelity factor~\cite{giovannettiQuantumLimitsDynamical2003,freyQuantumSpeedLimits2016}.
Here, a classical dynamical system with a finite average dissipation rate $\bar\Lambda$ cannot evolve between two states distinguishable by $\eta$ on a time shorter than $\Delta t_\eta \geq f(\eta)/\bar{\Lambda}$.
If $\Delta t_{\sfrac{1}{3}}$ is time for a given trajectory to reach the condition $\eta=\sfrac{1}{3}$, 
then $\Delta t_{\sfrac{1}{3}}\geq 1/2\bar\Lambda$, which is mathematically similar to the original Margolus-Levitin bound.

(3) Lastly, in the form $\bar\Lambda \Delta t_{\eta}\geq f(\eta)$, the inequality sets a finite lower bound on the accessible phase space volume (or the count of the distinct physical states) the system visits over the time interval.

Regardless of the form we choose, there are limiting cases of interest for the maximum speed set by this, and other measures of, dissipation.
The bound saturates if the net contraction rate is sufficiently small for higher powers of $\int dt \Lambda$ (cubic and beyond) to be negligible.
This condition is a result of using Pad\'e approximation, which ensures the power series of $f(\eta)$ matches the power series of the exponential up to second order: $e^x \approx \frac{2+x}{2-x} \approx 1 + x + x^2/2$.
Another limiting case is when the system is not dissipative $\bar \Lambda=0$. Over the course of a trajectory, the minimum time would then diverge because the states are indistinguishable, $\eta\to 1$.
However, at the other extreme, if the phase space contraction rate diverges or when the two states are perfectly distinguishable, $\eta\to 0$, then the minimum time is zero.
As confirmed by our numerical simulations, we expect actual dissipative dynamics systems to fall between these extremes where the trade-off between time and phase space contraction is significant.

This first speed limit on phase space contraction is well connected to existing results in nonequilibrium statistical mechanics.
As in quantum dynamics, the distinguishability criterion $\Gamma=\eta$ determines the numerical value for the minimum time or maximum speed.
Physical motivation for the criterion in Eq.~\ref{eq:survival} comes from the intuition that phase space contracts when a system is dissipative but also from the appearance of this criterion in \textit{deterministic} fluctuation theorems.
The compression factor $\Gamma$ in Eq.~\ref{eq:survival} is the probability that phase space contracts by an amount $A$ relative to the probability it expands by $-A$ over a time interval $\Delta t$ up to a two.
According to the transient fluctuation theorem~\cite{EvaCohMor1993,EvansSearles2002}
\begin{align}\label{eq:TFT}
  \frac{p_{\Delta t}(\bar{\Lambda} = +A)}{p_{\Delta t}(\bar{\Lambda}=-A)} = e^{+\Delta t A}=:+\sqrt{\eta},
\end{align}
increasing the system size or the time interval $\Delta t$
suppresses second law ``violations''~\footnote{The derivation of this fluctuation theorem assumes that the initial phase space distribution is time symmetric but otherwise arbitrary.}. 
From this perspective, the speed limit is the minimum time for the relative probability to reach a value $\eta$.
Like the fluctuation theorem, the speed limit applies to finite-size systems away from equilibrium. 
Recognizing this direct connection to fluctuation theorems makes it possible to derive a family of speed limits with physical observables, including the entropy production, heat, and transport coefficients.

\textit{Speed limit from the entropy production rate.---} The first step to deriving statistical mechanical speed limits is to instead consider an ensemble of systems.
Again, the probability density $\rho(\boldsymbol{x},t)$ of the ensemble evolves between two arbitrary distributions $\rho(\boldsymbol{x},t_0)$ and $\rho(\boldsymbol{x},t)$ over a time interval $\Delta t=t-t_0$ according to Liouville's equation.
At any moment of time in this interval, the Gibbs entropy rate is the ensemble average of the phase space contraction rate~\cite{Andrey1985,Ruelle1996,Ruelle1997}:
\begin{equation}\label{eq:gibbs-entropy-rate}
  \dot{S}_G/k_B = +\int d\bx \rho(\bx,t)\Lambda(\bx,t) = +\langle \Lambda\rangle.
\end{equation}
This rate is negative for strongly chaotic and dissipative dynamical systems with no vanishing Lyapunov exponents~\cite{Ruelle1996,Ruelle1997}, so a convention is to add an overall negative sign and interpret this as the entropy production rate in the environment~\cite{Klages2007}.

Another fluctuation theorem applies to these statistical ensembles of a system producing entropy away from equilibrium.
The integrated fluctuation theorem measures probability of second law violations of a small system in a time interval $\Delta t$~\cite{EvansSearles2002}. 
However, as our first speed limit here suggests, we can also use fluctuation theorems as a condition for the statistical distinguishability of states and derive corresponding speed limits.
Over a finite time interval, a system produces more entropy than it consumes by a factor 
$P_{\Delta t}(\bar\Lambda < 0)/P_{\Delta t}(\bar\Lambda >0) = \langle e^{-\Delta t\bar{\Lambda}}\rangle_{\bar{\Lambda} > 0}$.\\

The probabilities here are for any negative (positive) values of $\bar{\Lambda}$ in the time interval $\Delta t$ across the phase space.
Ensemble averaging Eq.~\ref{eq:survival},
\begin{align}\label{eq:sur-fac2}
\langle\eta\rangle := \langle \Gamma(t, t_0) \rangle =
\left\langle e^{-2\Delta t \bar \Lambda}\right\rangle_{\bar\Lambda >0},
\end{align}

the value of $\langle\eta\rangle$ defines a statistical distinguishability criterion based upon how much more likely a set of trajectory segments is to produce than consume entropy. 

From the exponential in Eq.~\ref{eq:sur-fac2}, Jensen's inequality gives:
$\langle \eta\rangle \geq e^{-2\Delta t \overline{S}_e/k_B}$

where $\overline{S}_e/ k_B = -\Delta t_\eta^{-1}\int dt \langle\Lambda\rangle \geq 0$ is the total entropy produced in the environment.
Applying the bilinear approximation as before
and defining the function $f(\langle\eta\rangle):=(1-\langle\eta\rangle)/(1+\langle\eta\rangle)$, we get
\begin{align}
  \Delta t_{\langle\eta\rangle} \overline{S}_e/k_B\geq f(\langle\eta\rangle),
\label{eq:entropy_speed_limit}
\end{align}
a trade-off between the Gibbs entropy rate and the time to transition between states.

This speed limit has three interpretations like the speed limit in Eq.~~\ref{eq:diss_time1}.
It also has the advantage that the ensemble average makes the bound independent of the initial condition for the underlying deterministic dynamics.
The bound instead applies to an ensemble transitioning between equilibrium or nonequilibrium states described by a probability density. 
For deterministically thermostatted systems, the Gibbs entropy rate is the thermodynamic entropy production rate~\cite{JeppsRondoni2010, HarPat2021}, so this inequality is a tradeoff between thermodynamic dissipation and time for any system transitioning between two statistical states distinguishable by $\langle\eta\rangle$. 
There is a corresponding bound for intensive observables, SM~\ref{SM-speed-limit}.

Stepping back, these two speed limits are closely connected and part of a much larger family of speed limits involving observables.
One immediate bound applies to mixing dynamical systems: The rate $\Lambda(t)$ is the sum over all instantaneous Lyapunov exponents, $\Lambda(t)=\sum_i^n\lambda_i(t)$, but we can simply restrict the sum to positive exponents to derive a speed limit set by the finite-time Kolmogorov Sinai entropy~\cite{GreenCGS13}, another signature of deterministic chaos $\Delta t_\eta h_{\scriptscriptstyle\textrm{KS}}\geq 1$.
However, looking at the many known fluctuation theorems~\cite{EvansSearles2002} suggests the approach here gives bounds on many physical quantities.
These follow from recognizing Equations~\ref{eq:diss_time1} and \ref{eq:entropy_speed_limit} as two fundamental realizations of the dissipation function, $\bar{\Lambda}$ and $\overline{\langle\Lambda\rangle}$.
Following the steps here for other expressions of the dissipation function then leads directly to speed limits on other observables~\cite{EvansSearles2002}, including the dissipative flux (SM~\ref{SM:diss-flux}), heat rate, and transport coefficients.
From a given fluctuation theorem, the minimum transition time (a quantity that is difficult to predict a priori) can be predicted by a physically measurable quantity.
To illustrate, we take a couple of important examples: the energy dissipated as heat and the electrical conductivity.

\textit{Speed limit from energy dissipated as heat.---} 

Another immediate physical observable subject to a speed limit is the energy exchanged as heat between a many-particle system and its environment.
For physical systems interacting with deterministic, time-reversible thermostats (e.g., Gaussian, Nos\'e-Hoover), the phase space contraction rate and the Gibbs entropy rate are a link between the properties of microscopic dynamics~\cite{JeppsRondoni2010, Hoovers2020} and nonequilibrium states~\cite{evans_morriss_2008}. 
Not only is the (negative) Gibbs entropy rate the thermodynamic entropy production rate~\cite{Hoover2016a, Hoover2016b, Ramshaw2017, Hoovers2020, Ramshaw2020} 

it is also directly related to the energy is dissipated as heat at a mean rate $\dot{Q}$, ${\dot{S}_e/k_B = \dot{S}_G/k_B= \dot{Q}/k_BT}$~\footnote{We use the convention that the heat rate is negative when the system dissipates energy to the surroundings.}. 

Using this identity, Eq.~\ref{eq:entropy_speed_limit} becomes
\begin{align}\label{eq:speed-heat-prod}
  \Delta t_\eta \bar{Q} \geq k_BT,
\end{align}
an energy-time tradeoff with the temperature setting a lower bound on the minimum time for the system to exchange an amount of energy as heat with its environment, $\bar Q/k_BT = -\Delta t_\eta^{-1}\int \langle\Lambda\rangle dt$.
A higher heat exchange rate relative to the typical energy fluctuations, $k_BT$, will lower the minimum transition time between two statistical states.

\begin{figure}[t!]
\hspace*{-0.1cm}\includegraphics[width=1\columnwidth]{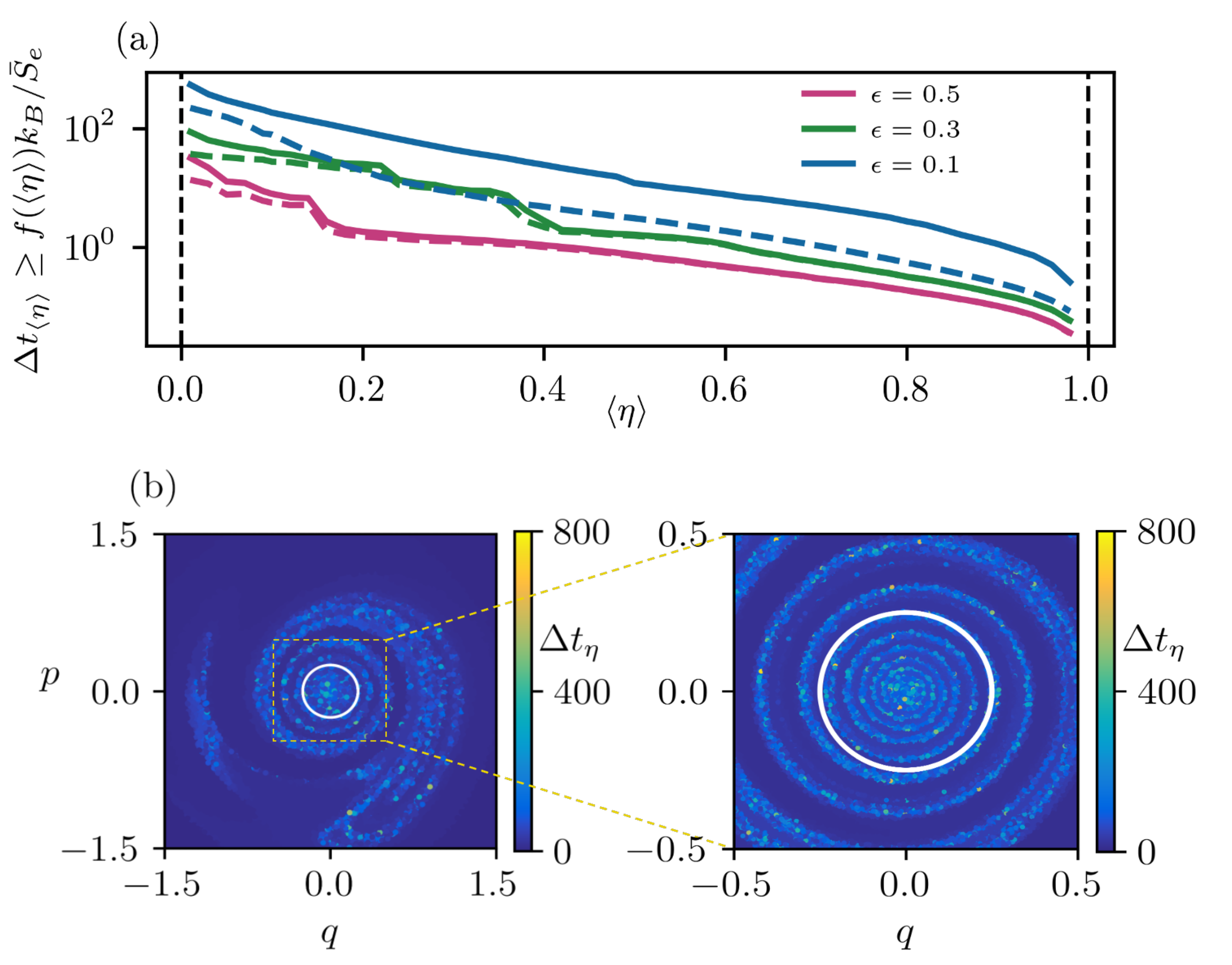}
\vspace{-0.2cm}
\caption{\label{fig:0532-model} \textit{Increasing the entropy rate suppresses the transition time in a thermally-conducting oscillator.---} (a) The time elapsed for the net dissipation as measured by $\langle \Gamma\rangle$ to reach the threshold $0 < \eta< 1$ from numerical simulations of the 0532 model, $(\alpha, \beta)=(0.05, 0.32)$.
Data are for three representative values of the temperature gradient parameter $\epsilon \in \{0.1, 0.3, 0.5\}$ (solid blue, green, and pink lines).
The minimum time for the ensemble to reach those $\eta$ thresholds (dashed lines) set by the entropy produced in the environment $f(\langle\eta\rangle)k_B/\bar{S}_e$ 
with $f(\langle\eta\rangle)=(1-\langle\eta\rangle)/(1+\langle\eta\rangle)$.
Averages are over $10^5$ initial conditions $(q_0, p_0, \xi_0 = 0)$ randomly chosen from the uniform distribution in the interval $\{-1.5, 1.5\}$ for each value of $\eta$.
(b) In the $\zeta = 0$ phase plane $10^6$ initial conditions show dissipative structures: the time $t_{\eta}$ to reach $\eta=\sfrac{1}{3}$ is longer for trajectories initiated from points on logarithmic spirals emanating from the fixed point at $(0,0)$. An orbit of the harmonic oscillator is shown for comparison (white circle).}
\end{figure}

\textit{Speed limit from transport coefficients.---}

As another example, Eq.~\ref{eq:entropy_speed_limit} also extends to transport coefficients. For systems interacting with Gaussian and Nos\'e-Hoover thermostats, the electrical conductivity, diffusion, and viscosity are functions of $\langle\Lambda\rangle$~\cite{Klages2000, Klages2007}. 
Consider the Lorentz gas in which charged particles (neglecting Coulomb interactions) move in a two-dimensional array of fixed hard disk scatterers. Particles are also driven by an external electric field $\mathbf{E}$ and thermostatted by a reservoir at temperature $T$~\cite{HenkDorfman1995}. 
The electrical conductivity is $\sigma (t) = T/\mathbf{|E|^2}\overline{\langle\Lambda(t)\rangle}$.
From Eq.~\ref{eq:entropy_speed_limit}, the electric field and temperature set the lower bound: $\Delta t_\eta\sigma(t) \geq T/2\mathbf{|E|^2}$ for $\eta = 1/3$.
Intuitively, stronger electric fields or higher conductivity can expedite the transition between states, while higher temperatures will delay the transition.
Similar intuition comes from the simpler case of the harmonic oscillator with friction coefficient $\kappa$ and mass $m$: for $\eta = 1/3$, the speed limit $\Delta t_\eta\kappa/m\geq 1/2$ suggests that decreasing the friction or increasing the mass will lower the minimum the time to dissipate a given amount of energy.

\textit{Numerical simulations of a thermally-conducting oscillator.---} As a numerical test case, we analyze the dynamics of a one-dimensional oscillator interacting with a Nos\'e Hoover thermostat~\cite{Patra2016}, SM~\ref{SM:0532-model}.
The position and momentum $(q, p)$ of the oscillator are thermostatted at temperature $T$ by the linear friction force $-\zeta p$.
For any nonzero with friction coefficient $\zeta$,  
the phase space contraction rate is $\Lambda(t) = - \zeta\left(\alpha + 3\beta p^2/T\right)$. 
A negative (positive) $\Lambda$ indicates heat lost to (gained from) the thermostat.
To simulate the ``0532 model'', we choose $(\alpha, \beta) = (0.05, 0.32)$, which is ergodic with a Gibbsian phase space distribution~\cite{Hoover2016b} at thermal equilibrium.
More importantly here, adding a local temperature gradient with a profile $T = 1 + \epsilon\tanh(q)$ and parameter $0<\epsilon< 1$ drives the system away from equilibrium.
To compute the speed limit from the entropy production, we chose three representative values of $\epsilon \in \{0.1, 0.3, 0.5\}$.

Figure~\ref{fig:0532-model} shows the transition time and the lower bound $\overline{\langle\Lambda\rangle}$.
These data are averaged over $10^5$ initial conditions $(q_0,p_0, 0.0)$ randomly chosen in the interval $[-1.5, 1.5]$ on the $\zeta = 0$ plane for 50 equally spaced values of $\eta$ between 0 and 1.
They show that as the ensemble relaxes to the nonequilibrium steady state attractor, the time intervals in which trajectories first satisfy the $\eta$ threshold is lower bounded by $f(\eta)/\bar S_e$.
Increasing the nonequilibrium drive from $\epsilon=0.1$ to $0.5$ increases the entropy produced and decreases the transition time. 

Across the $(q,p)$ phase space, the speed limit resolves structures that are dissipative on a particular timescale.
For $\eta=\sfrac{1}{3}$, the $\zeta = 0$ plane in the second panel of Fig.~\ref{fig:0532-model}.
In this case, this speed limit is: $\Delta t\bar s_i \geq \frac{1}{2}$.
For improved resolution, we randomly sample $10^6$ initial points and color them according to their $\Delta t$ at which the corresponding trajectories take to attain the $\eta = \sfrac{1}{3}$.
Those regions on the $\zeta = 0$ plane with higher values of $\Delta t_\eta$ are correlated with the smaller values of $\bar{\Lambda}$, satisfying our speed limit on dissipation in Eq.~\ref{eq:diss_time1} in each case.
These phase points dissipating less appear as logarithmic spirals approaching the fixed point of the system at the origin.
These structures show that there are some regions of phase space that are slower and others that are faster to transition between states, up to but not exceeding the maximum speed set by phase space contraction rate.

The speed limits on dissipation here are statistical and mechanical, applying to many physical observables, and quantifying the natural tension between dissipation and the time to produce it during irreversible processes.
They results bridge the deterministic, potentially chaotic, dynamics of physical systems, avoiding any hypotheses about chaos and instead making measures of chaos explicit through the contraction of phase space.
These bounds are a quantitative statement of how the production of entropy, expulsion of energy as heat, or transport of matter will decrease the minimum time to evolve between two distinguishable physical states.

\textit{Acknowledgments.---}
This material is based upon work supported by the National Science Foundation under Grant No. 2124510. 

\bibliography{references.bib}

\clearpage
\setcounter{figure}{0}
\renewcommand{\thefigure}{SM\arabic{figure}}
\title{Supplemental Material: Maximum speed of dissipation}
\maketitle
\setcounter{equation}{0}
\setcounter{section}{0}
\section{Equation of motion for the trace of the density matrix}\label{SM:EOM-xi}
A perturbation vector in the tangent space of a classical trajectory evolves according to:
\begin{equation}
d_t\ket{\delta \bx_i(t)} = \stability[\bx(t)]\ket{\delta \bx_i(t)}.
\label{sm-equ:EOM_per_ket2}
\end{equation}
For many-body systems, these vectors are perturbations to positions and momenta $\delta\bx_i = (\delta\bq^i, \delta\bp_i)$ .
The stability matrix, $\stability:=\stability[\bx(t)] = \grad\boldsymbol{F}$ generating these dynamics has the elements $(\stability)^i_{j}=\partial \dot{x}^i(t)/\partial x^j(t)$.
Dirac's notation here represents a finite-dimensional vector (its transpose) with the ket (bra): $\ket{\delta\bx(t)}:= [\delta x^1(t), \delta x^2(t), \ldots, \delta x^n(t)]^\top\in T\mathcal{M}$. Solving the equation of motion~\cite{PikovskyP16}:
\begin{align}\label{equ:per_ket}
\ket{\delta \bx_i(t)}
&= \mathcal{T}_+e^{\int_{t_0}^{t}\stability(t')\,dt'}\ket{\delta \bx_i(t_0)},
\end{align}
gives the perturbation vector at a time $t$ (with the time ordering operator $\mathcal{T}_+$ because stability matrices at two times do not generally commute).

To span the volume $\delta \mathcal{V}$, we can choose a complete set of $n$ linear independent unit vectors $\{\ket{\delta\bphi_i}$\} and
normalize them $\delta\bphi_i = \delta\bx_i/\|\delta\bx_i\|$ so that the outer product defines a classical density matrix, $\bxi(t) = \sum_{i=1}^{n}\dyad{\delta\bphi_i}{\delta\bphi_i}$. As the basis vectors form  complete set of independent vectors, the determinant of $\bxi$ gives the volume $\delta \mathcal{V}$ span by them in the phase space.

In our previous work, see Ref.~\cite{DasGreen2022}, we have shown that the determinant of $\bxi$ denoted by $|\bxi|$ has the following equation of motion:
\begin{align}
	d_t \ln |\bxi| = 2\Tr\stability = 2\Lambda.
\end{align}
This gives rate of change of phase space volume under any non-conservative dynamics.
The evolution of its trace $\Tr\bxi$ is governed by:
\begin{align}\label{sm-eq:EOM-tr-xi}
	d_t \ln  \Tr\bxi = 2\Lambda/n.
\end{align} 
For a single tangent vector $\ket{\delta \bx_i(t)}$, the trace $\Tr \bxi_i(t) = \bra{\delta \bx_i(t)}\ket{\delta \bx_i(t)}$ gives the rate at which the perturbation represented by the vector $\ket{\delta \bx_i(t)}$ evolves in time.

The solution to Eq.~\ref{sm-eq:EOM-tr-xi} reads:
\begin{align}
\Tr\bxi(t) = \Tr\bxi(t_0)e^{2n^{-1}\int_{t_0}^{t}\Lambda(t') dt'}.
\end{align}

We use this solution to define the compression factor $\Gamma(t,t_0) = \Tr\bxi(t_0)/\Tr\bxi(t_0)$ which leads to the dissipation-time bound in the main text.

\section{Geometric interpretation of the density matrix}\label{SM:geometric-xi}
\begin{figure}[!t!]
	\centering
	\begin{tikzpicture}[thick, >=stealth]
	\hspace*{-2.9cm}

	\begin{scope}[rotate=120, xscale=0.45, yscale=1.3, shift={(0.0,0.0)}]
	\coordinate (O) at (0,0);
	\shade[ball color=cyan!50!black,opacity=0.6] (0,0) coordinate(Hp) circle (1) ;
	\end{scope}
	\draw [->,>=stealth,thick, red, line width=1pt] (0,0) -- (1.1,0.6) node [at end, right] {\small $\ket{\delta \bphi_1}$};
	\draw [->,>=stealth,thick,  red, line width=1pt] (0,0,0) -- (-0.36/1.5,0.4,0.0) node [at end, above] {\small $\ket{\delta \bphi_2}$};
	\draw [->,>=stealth,thick,  red, line width=1pt] (0,0,0) -- (0.16,-0.1,1.48) node [at end, below] {\small $\ket{\delta \bphi_3}$};
	\tdplotsetmaincoords{180}{0}
	\tdplotsetrotatedcoords{0}{0}{0}
	\node[] (O) at (0,0,0){};
	\hspace{2.5cm}
		\draw [<->,>=stealth,thick, line width=2pt] (-0.5,0,0.0) -- (0.5,0.0,0.0) node [at end, left]{};
		
	\hspace{2.2cm}	
		\draw [->,>=stealth,thick,blue, line width=1pt] (0,0,0) -- (1.0,.18,0.05) node [at end, right] {\small $\propto (\Tr \bxi)^{1/2}$};
			\tdplotsetmaincoords{180}{0}
	\tdplotsetrotatedcoords{0}{0}{0}
	\node[] (O) at (0,0,0){};
	\begin{scope}[rotate=0, xscale=1.0, yscale=1.0, shift={(0.0,0.0)}]
	\coordinate (O) at (0,0);
	\shade[ball color=cyan!50!black,opacity=0.6] (0,0) coordinate(Hp) circle (1) ;
	\node at (-2.1,-0.8) {\color{black}$\delta \mathcal{V} = |\bxi|^{1/2}$};
	\end{scope}
	\end{tikzpicture}
	\caption{A compressed phase space volume at any time $t$ can be mapped onto a hypersphere in the phase space using the classical density matrix framework. This schematic shows a contracted phase volume  spanned by a complete set of basis $\{\ket{\delta \bphi_1}, \ket{\delta \bphi_2}, \ket{\delta \bphi_3}\}$ in a three dimensional phase space with an arbitrary shape. Its volume  $\delta \mathcal{V}$ is given by the square root of determinant of the density matrix $|\bxi(t)|^{1/2}$, where  $\bxi(t) = \sum_{i=1}^3\dyad{\delta \bphi_i}{\delta \bphi_i(t)}$. This contracted phase space is equivalent to a sphere of the same volume $\delta \mathcal{V}$ and radius proportional to $\Tr\bxi(t)^{1/2}$ at any time $t$.  This geometric picture holds  for an $n$-dimensional phase space. \label{sm-fig-volume}}  
	\end{figure}
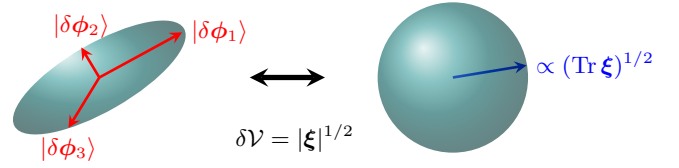
The dynamic of a unit vector in the tangent space can be normalized at every time. The equation of motion of unit tangent vector $\ket{\delta \yu_i} :=\ket{\delta \bx_i(t)}/\|\delta \bx_i(t)\|$ is given by:
\begin{align}
	d_t\ket{\delta \yu_i(t)}:= \stability\ket{\delta \yu_i(t)} - r_i(t)\ket{\delta \yu_i},
\end{align}
where $r_i(t) = \bra{\delta \yu_i(t)}\stability(\bx)\ket{\delta \yu_i(t)}$ can be identified as the instantaneous Lyapunov exponent.

For this normalized dynamics, we can construct a density matrix $\brho_i \ \ket{\delta \yu_i(t)}\bra{\delta \yu_i(t)}$ which accounts for the normalized evolution of $\ket{\delta \yu_i}(t)$ with time. The trace of $\brho_i$ remains 1 for all times.

As before, we consider the complete set of linearly independent unit vectors ${\ket{\delta \bphi_i}}$ to use their outer product to construct density matrix $\brho=n^{-1}\sum_{i=1}^{n}\dyad{\delta\bphi_i}{\delta\bphi_i}$. Due to the normalized evolution of $\ket{\delta \bphi_i}$ for all times, $\Tr \brho = 1$ also remains time invariant. The matrices $\brho$ and $\bxi$ and their determinants are related as follows:
\begin{align}
	\brho &= \frac{\bxi}{\Tr\bxi}\\
	\implies |\brho|&= \frac{|\bxi|}{(\Tr \bxi)^n},
\end{align}
at any time $t$ in an $n$-dimensional phase space.

The determinants $|\bxi(t)|$ and $|\brho (t)|$ give different phase space volumes at time $t$. But, since $\brho$ generates normalized evolution, the volume it represents is a constant of motion. This observation leads to the generalized Liouville's theorem and equation we have proposed in Ref.\cite{DasGreen2022}. 

This shows that $|\bxi(t)| \propto (\Tr\bxi(t))^n$, see Fig.~\ref{sm-fig-volume}. For a geometric interpretation, imagine a hypersphere in the phase at time $t$ attached to the given trajectory. The volume of this hypersphere is given by $|\bxi(t)|^{1/2}$ and radius $(\Tr\bxi(t))^{1/2}$. This $n-$sphere represents a contracted phase space volume in a dissipative system at time $t$. We have thus used the density matrix formalism to map the compressed phase space volume of an arbitrary shape at time $t$  to a hypersphere.

In the next section, we use the observation  $|\bxi(t)| \propto (\Tr\bxi(t))^n$ to obtain intensive bounds. These are another set of speed limit on dissipation \section{Derivation of the speed limit on dissipation (System-size intensive)}\label{SM-speed-limit} 
Consider the compression factor,
\begin{align}
	\Gamma(t, t_0) = e^{2\int_{t_i}^{t_f}dt\Lambda/n}.
\end{align}

Let $\Delta t_\eta$ is the time interval in which $\Gamma$ reduces by a factor of $0\leq\eta<1$. This leads to
\begin{align}
\eta = e^{2\int_{t_i}^{t_f}dt\Lambda/n}.
\end{align}
Applying a bilinear approximation of $e^{2x}$ for $x\leq 0$, we obtain
\begin{align}
	\eta \geq \frac{1 - \Delta t_\eta \bar{\Lambda}/n}{1 + \Delta t _\eta\bar{\Lambda}/n},
\end{align}
where $\bar{\Lambda} = -\Delta t_\eta^{-1}\int_{t_i}^{t_f}dt\Lambda$. Rearranging the inequality, 
\begin{align}
	\Delta t_\eta\bar{\Lambda}/n \geq \frac{1 - \eta}{1 + \eta};\quad (\bar{\Lambda} > 0,0 < \eta\leq 1).
\end{align}

This inequality is the system-size intensive version of Eq.~2 (main text). We also give the entropy speed limit, which is obtained following the steps which yield Eq.~6 (main text):
\begin{align}
  \Delta t_{\langle\eta\rangle} \overline{S}_e/nk_B\geq \frac{1 - \langle\eta\rangle}{1 + \langle\eta\rangle};\quad (0 < \langle\eta\rangle\leq 1),
\end{align}
where $\overline{S}_e/ k_B = -\Delta t_\eta^{-1}\int dt \langle\Lambda\rangle \geq 0$ is the total entropy produced in the environment.

A similar set of relations can be found if we consider the expansion of phase volume  $\Lambda \geq 0$ i.e. $\bar \Lambda \leq 0$.  
In dissipative systems, there may be time intervals in which $\Gamma$ increases implying that $\eta \geq 1$. The bound in Eq.~2 (main text) then takes the following form:
\begin{align}\label{eq:speed-limit-2}
\quad -\Delta t_\eta \bar{\Lambda}/n \geq \frac{\eta - 1}{\eta + 1}; \quad (\bar{\Lambda} < 0, \eta\geq 1).
\end{align}
This inequality holds for small net expansion $0\leq -\Delta t\bar\Lambda = \int dt \Lambda< 1$. The ensemble average version of the bound in Eq.~\ref{eq:speed-limit-2} follows as before.

\section{Speed limit from dissipative flux}\label{SM:diss-flux}
In this section, we will show a possible speed limit from the dissipative flux $J$ in isothermal systems. A typical example is a set of $N$-particle in the presence of an external dissipative field $F_e$. Following Ref.~\cite{EvansSearles2002}, we give the transient fluctuation theorem for isokinetic dynamics:
\begin{align}
\frac{p(\bar{J} = A)}{p(\bar{J} = -A)} = e^{AtF_eV/(k_BT)},
\end{align}
where $\bar J = -\Delta t^{-1}\int dt' J$ defines the time average of the dissipative flux $J$ and $V$ represents the volume of the system of interest. This theorem holds for all times for the isokinetic ensemble when all initial phases are sampled from an equilibrium isokinetic ensemble.

We define the compression factor $\eta =  e^{2AtF_e V/(k_BT)}$ in accordance with our formulation described in the main text and follow Eq.~2 (main text) to write the following bound :
\begin{align}
\Delta t_\eta \bar JF_e V\geq f(\eta)k_BT.
\end{align}
The speed limit from the rate of the ensemble average $\langle J\rangle$ has a form similar to Eq~6 (main text):
\begin{align}
\Delta t_\eta \overline{\langle J\rangle} F_e V\geq f(\langle\eta\rangle)k_BT.
\end{align}
We have derived these bounds with assumption that $F_e$ is a constant. Similar speed limits can be derived from all physical observables which follow the fluctuation theorem. For a list, see table 4.1 in Ref.~\cite{EvansSearles2002}.

\section{Singly-thermostated harmonic oscillator: The 0532 model}\label{SM:0532-model}
This recently introduced system belongs to the family of Nos\'e-Hoover thermostats. It is one of the simplest example to investigate thermodynamic dissipation in the thermostatted dynamics of the one-dimensional harmonic oscillator. The system is given by oscillator coordinate $q$, velocity $p$, and friction coefficient $\zeta$ at a thermostat temperature $T$:
\begin{equation}
\begin{aligned}\label{sm-eq:thermostat_0532}
\dot{q} &= p, \quad \dot{p} = -q - \zeta\left(\alpha p + \frac{\beta p^3}{T}\right),\nonumber\\
\dot{\zeta} &= \alpha\left(\frac{p^2}{T} - 1 \right) + \beta \left(\frac{p^4}{T^2} -  \frac{3p^2}{T}\right),
\end{aligned}
\end{equation}
with parameters, $\alpha$ and $\beta$, set to $0.05$ and $0.32$. These equations impose a \textit{weak} control of the second and fourth moments of $p$. The control variable $\zeta$ is responsible for stabilizing the kinetic energy of the system.

The system is deterministic, time-reversible, and has been shown to be ergodic through extensive numerical tests. At equilibrium~\cite{Patra2016, Hoover2016b} the model reproduces the Gibbs distribution $f(q,p,\zeta)\propto e^{-q^2/2T}e^{-p^2/2T}e^{-\zeta^2/2T}$ in units with $k_B=1$. Moreover, in these ergodic thermostats, the time-averaged heat current $\overline{p^2}/2$ measured directly from the model has been found to be in excellent agreement with the Green-Kubo linear-response theory~\cite{Patra2016}.

To study thermodynamic entropy production, a localized temperature gradient can be introduced using a smooth temperature profile $T = 1 + \epsilon \tanh(q)$ with $0\leq \epsilon <1$. The localized temperature within the system thus varies between $1-\epsilon$ and $1+\epsilon$:
\begin{align}
	T(-\infty) = 1-\epsilon \leq T(q) \leq 1 + \epsilon = T(+\infty).
\end{align}
For a nonzero value of $\epsilon$, the system is out of equilibrium. The temperature gradient $dT/dq = \epsilon/\cosh^2 (q)$ generates a heat transfer from (to) the system to (from) the thermostat making the dynamics dissipative and irreversible.\\

\begin{figure}[t!] 
\hspace*{-0.1cm}\includegraphics[width=1\columnwidth]{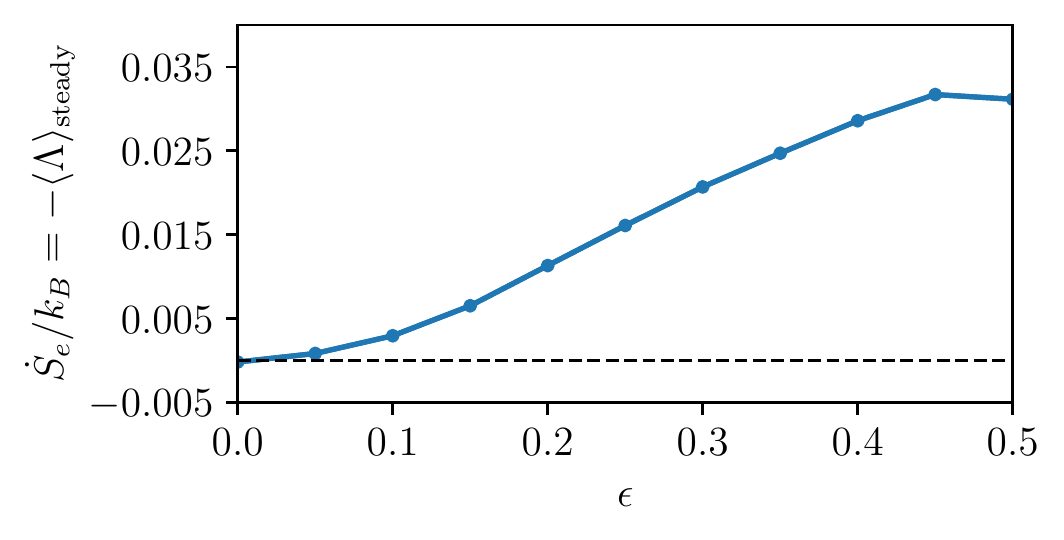}
\vspace{-0.2cm}
\caption{\label{fig:avg-entropy-eps} \textit{The NESS  value of the thermodynamic entropy production in the 0532 thermostat with increasing temperature gradient parameter $\epsilon$}. Averages are over $10^5$ initial conditions $(q_0, p_0, \xi_0 = 0)$ randomly chosen from the uniform distribution in the interval $\{-1.5, 1.5\}$ for 12 different values of $\eta$ in the interval $[0.0, 0.5]$. For time averaging, the $\Lambda$ values were taken after $10^6$ iterations for each trajectory. The long time average of $\Lambda$ followed by its large ensemble average gives the NESS value of $\langle\Lambda\rangle$ for each $\epsilon$.}
\end{figure}

The phase space volume contraction rate $\Lambda$ is given by:
\begin{align}\label{sm-eq:lambda}
	\Lambda = \frac{\partial \dot{q}}{\partial q}+\frac{\partial \dot{p}}{\partial p}+\frac{\partial \dot{\zeta}}{\partial \zeta} = - \zeta\left(\alpha + \frac{3\beta p^2}{T}\right).
\end{align}
At any time $t$, the instantaneous rate of heat loss/gain $\dot h$ by the system can be obtained as follows:
\begin{align}\label{sm-eq:energy}
	\dot{h} = q\dot{q} + p\dot{p} = - \zeta \left(\alpha p^2 + \frac{\beta p^4}{T}\right).
\end{align}

To fix the kinetic energy, the time average of the square of thermostatted variable, $\bar{\zeta}^2$ needs to remain a constant. This means that $\dot{\zeta}$ must vanish at all times and we can write:
\begin{align}
	\overline{\zeta\dot{\zeta}}  = 
	\overline{\zeta\alpha\left(\frac{p^2}{T} - 1 \right) + \zeta\beta \left(\frac{p^4}{T^2} -  \frac{3p^2}{T}\right)} & = 0. \nonumber
\end{align}
We can rearrange this equation to find:
\begin{align}
	\overline{- \zeta\left(\alpha + \frac{3\beta p^2}{T}\right)} &= \overline{- \frac{\zeta}{T} \left(\alpha p^2 + \frac{\beta p^4}{T}\right)}\nonumber\\
	\implies\overline{\Lambda} &= \overline{\frac{\dot h}{T}}, \label{sm-eq:heat-rate-lambda}
\end{align}
from Eqs.~\ref{sm-eq:lambda} and ~\ref{sm-eq:energy}. We now take ensemble average both sides of Eq.~\ref{sm-eq:heat-rate-lambda} and identify $\langle \Lambda\rangle$ as the Gibbs entropy rate $\dot{S}_G$ to find the following identities:
\begin{align}
	\frac{\dot{S}_e}{k_B} = \frac{\dot{S}_G}{k_B} = -\langle\Lambda\rangle = \frac{\dot{\mathcal{Q}}}{T},
\end{align}
where $\dot{\mathcal{Q}}= \langle \dot h\rangle$ is the rate of average exchange of heat of the system with its environment. These ensemble average can be replaced by time averages as the system is ergodic. The Gibbs entropy rate gives the thermodynamic entropy production rate in thermostatted Hamiltonian systems of Nos\'e-Hoover and Gaussian types. 

The non-equilibrium steady state (NESS) value of $\dot S_e/k_B$ increases with $\epsilon$ in the interval $[0.0, 0.5]$ as shown in Fig.~\ref{fig:avg-entropy-eps}.

\vspace{1cm}
\onecolumngrid
\noindent\rule{\textwidth}{1pt}

\end{document}